\documentclass{IEEEtran}
\usepackage{cite}
\usepackage{amsmath,amssymb,amsfonts}
\usepackage{algorithmic}
\usepackage{graphicx}
\usepackage{textcomp}
\usepackage{balance}
\usepackage{braket}
\usepackage{color}
\usepackage{dcolumn}
\usepackage{bm}
\usepackage[utf8]{inputenc}
\usepackage[mathscr]{eucal}

\usepackage{mathtools}
\usepackage{mathrsfs}  
\usepackage{subcaption}

\usepackage{comment}
\DeclareMathOperator{\E}{\mathbb{E}}

\begin{document}

\title{A Hybrid Quantum-Classical \\Generative Adversarial Network \\for Near-Term Quantum Processors}
\author{Albha O'Dwyer Boyle and Reza Nikandish
\thanks{A. O'Dwyer Boyle was with the School of Electrical and Electronic Engineering, University College Dublin, Ireland.}
\thanks{R. Nikandish is with the School of Electrical and Electronic Engineering, University College Dublin, Ireland, and also with the Center for Quantum Engineering, Science, and Technology, University College Dublin, Ireland (e-mail: reza.nikandish@ieee.org).}}

\maketitle

\begin{abstract}
In this article, we present a hybrid quantum-classical generative adversarial network (GAN) for near-term quantum processors. The hybrid GAN comprises a generator and a discriminator quantum neural network (QNN). The generator network is realized using an angle encoding quantum circuit and a variational quantum ansatz. The discriminator network is realized using multi-stage trainable encoding quantum circuits. A \textit{modular design approach} is proposed for the QNNs which enables control on their \textit{depth} to compromise between accuracy and circuit complexity. Gradient of the loss functions for the generator and discriminator networks are derived using the same quantum circuits used for their implementation. This prevents the need for extra quantum circuits or auxiliary qubits. 
The quantum simulations are performed using the IBM Qiskit open-source software development kit (SDK), while the training of the hybrid quantum-classical GAN is conducted using the mini-batch stochastic gradient descent (SGD) optimization on a classic computer. The hybrid quantum-classical GAN is implemented using a two-qubit system with different discriminator network structures.
The hybrid GAN realized using a five-stage discriminator network, comprises 63 quantum gates and 31 trainable parameters, and achieves the Kullback-Leibler (KL) and the Jensen-Shannon (JS) divergence scores of 0.39 and 0.52, respectively, for similarity between the real and generated data distributions.

\end{abstract}


\maketitle

\section{Introduction}
Quantum computing can potentially lead to an unparalleled boost in computational power by using prominent features of quantum physics including superposition, entanglement, interference, and parallelism. This computational advantage can lead to revolutions in many emerging technologies which are hindered by current computing limitations in the software and hardware levels. The current quest for a practical quantum computer, when successful, can provide significant benefits to several applications including machine learning \cite{nature_machine_learning, intro_quantum_ML, farhi2018classification, schuld2014quest, abbas2021power, beer2020}, autonomous driving \cite{Luckow2021}, healthcare and drug discovery \cite{cao2018}, cryptography and secure communications \cite{gisin2002, Bernstein2017}, teleportation \cite{ren2017}, and internet \cite{kimble2008}.

Since the conception of quantum computing in 1980s \cite{feynman82}, this field has experienced several landmark developments to prove its promising capabilities and attract recent interests. The most notable inventions in the algorithm level include the Shor's algorithm for prime number factorization and discrete logarithm on a quantum computer \cite{Shor1994}, the Grover's algorithm for efficient search in unstructured large databases \cite{Grover1996}, the Deutsch–Jozsa algorithm for classification of a binary function as constant or balanced \cite{Deutsch_Jozsa}, and the HHL algorithm for solving linear systems of equations \cite{HHL}.

Quantum computers can outperform classic computers by the virtue of using quantum algorithms \cite{Montanaro2016}. Quantum algorithms are traditionally developed for \textit{long-term} fault-tolerant quantum computers with a large number of nearly perfect qubits (e.g., over 1 million). The quality of such algorithms is evaluated by their asymptotic computational complexity, i.e., how the quantum algorithm can perform a given computational task faster than a classic algorithm. The impressive speed advantage predicted by quantum algorithms, e.g., quadratic speedup for Grover's search algorithm, $\mathcal{O}(\sqrt{N})$, and exponential speedup for quantum support vector machine (SVM), $\mathcal{O}(\log{N})$, has been a driving force in the recent extensive efforts on quantum computers \cite{nature_machine_learning}.

The \textit{near-term} quantum computers, aka Noisy Intermediate-Scale Quantum (NISQ) computers \cite{preskill_nisq}, are realized using \textit{imperfect} and \textit{limited} number of qubits (e.g., 10--100). The quantum circuits realized using these noisy qubits should have a limited depth to avoid the loss of quantum state by the quantum decoherence. These quantum computers cannot provide computational resources required to leverage the fault-tolerant quantum algorithms and, as a result, can outperform classic computers only in a few computational tasks \cite{nature_machine_learning, preskill_nisq}. The near-term quantum computers are realized to demonstrate quantum advantages for some specific computational tasks \cite{arute2019quantum, Madsen2022}.
A quantum algorithm should account for the inherent limitations of the NISQ computers, e.g., limited number of qubits, decoherence of qubits, limited depth of quantum circuits, and poor connectivity of qubits, to be able to achieve a quantum advantage. Variational quantum algorithms (VQAs) have emerged as one of the, if not the, most effective approaches in the near-term quantum computing era \cite{cerezo2021variational, Tacchino2021, cerezo2022}. 
Quantum machine learning can significantly benefit from the VQAs which can also be interpreted as quantum neural networks (QNNs) \cite{farhi2018classification, schuld2014quest, nature_machine_learning, intro_quantum_ML, abbas2021power, beer2020}. The quantum circuit parameters can be trained by an optimizer using a predefined loss function running on a classic computer. Nevertheless, there are several challenges in trainability, efficiency, and accuracy of the VQAs which should be addressed through innovative solutions \cite{caro2022, cerezo2022}.

Most of quantum machine learning models are inspired by a classic model. Generative adversarial network (GAN), proposed by Goodfellow in 2014 \cite{GAN}, is a powerful tool for classical machine learning, in which a generator model captures the data distribution and a discriminator model maximizes the probability of assigning the correct true/fake label to data. Quantum GAN was proposed in 2018 for accelerating the classical GAN \cite{QGAN_init, lloyd2018quantum}. It is shown that in a fully quantum implementation, i.e., when the generator and discriminator are realized on a quantum processor and the data is represented on high-dimensional spaces, the Quantum GAN can achieve an \textit{exponential} advantage over the classic GAN \cite{lloyd2018quantum}.

This pioneering work on Quantum GAN were followed by a number of other developments presented in the literature \cite{zoufal2019quantum, romero2021variational, huang2021quantum, hu2019, niu2022entangling}. In \cite{zoufal2019quantum}, the efficient loading of statistical data into quantum states is achieved using a hybrid quantum-classical GAN. In \cite{romero2021variational}, a hybrid quantum-classical architecture is proposed to model continuous classical probability distributions. An experimental implementation of the Quantum GAN using a programmable superconducting processor is presented in \cite{huang2021quantum}, in which the generator and discriminator are realized as multi-qubit QNNs, and the generator could replicate an arbitrary mixed state with 99.9\% fidelity. In another experimental Quantum GAN, also implemented on a superconducting quantum processor \cite{hu2019}, an average fidelity of 98.8\% is achieved for both of the pure and mixed quantum states. A Quantum GAN architecture is proposed in \cite{niu2022entangling} which improves convergence of the minimax optimization problem by performing entangling operations between the generator output and true quantum data. These developments showcase some potentials of the Quantum GAN as a promising near-term quantum algorithm, while still there are many open challenges to the efficient realization, training, and application of the GAN in the quantum domain. 

In this article, we present a hybrid quantum-classical GAN for the near-term quantum processors. The generator network comprises a quantum encoding circuit and a variational quantum ansatz. The discriminator network is realized using a modular design approach which enables control on the \textit{depth} of quantum circuits to mitigate impacts of the imperfect \textit{low-fidelity qubits} in the near-term quantum processors.

The article is structured as follows. In Section II, architecture of the hybrid quantum-classical GAN, the gradient evaluation for quantum circuits, and the Barren Plateaus issue are presented. The quantum neural architecture search approach is discussed in Section III. In Section IV, training of the variational quantum ansatzes in the hybrid GAN is presented. In Section V, results of the implemented hybrid GAN are presented and discussed. Finally, concluding remarks are summarized in Section VI.

\section{Hybrid Quantum-Classical GAN}

\subsection{Hybrid GAN Architecture}

\begin{figure*}[!t]
    \centering
    \includegraphics[width = 1.7\columnwidth]{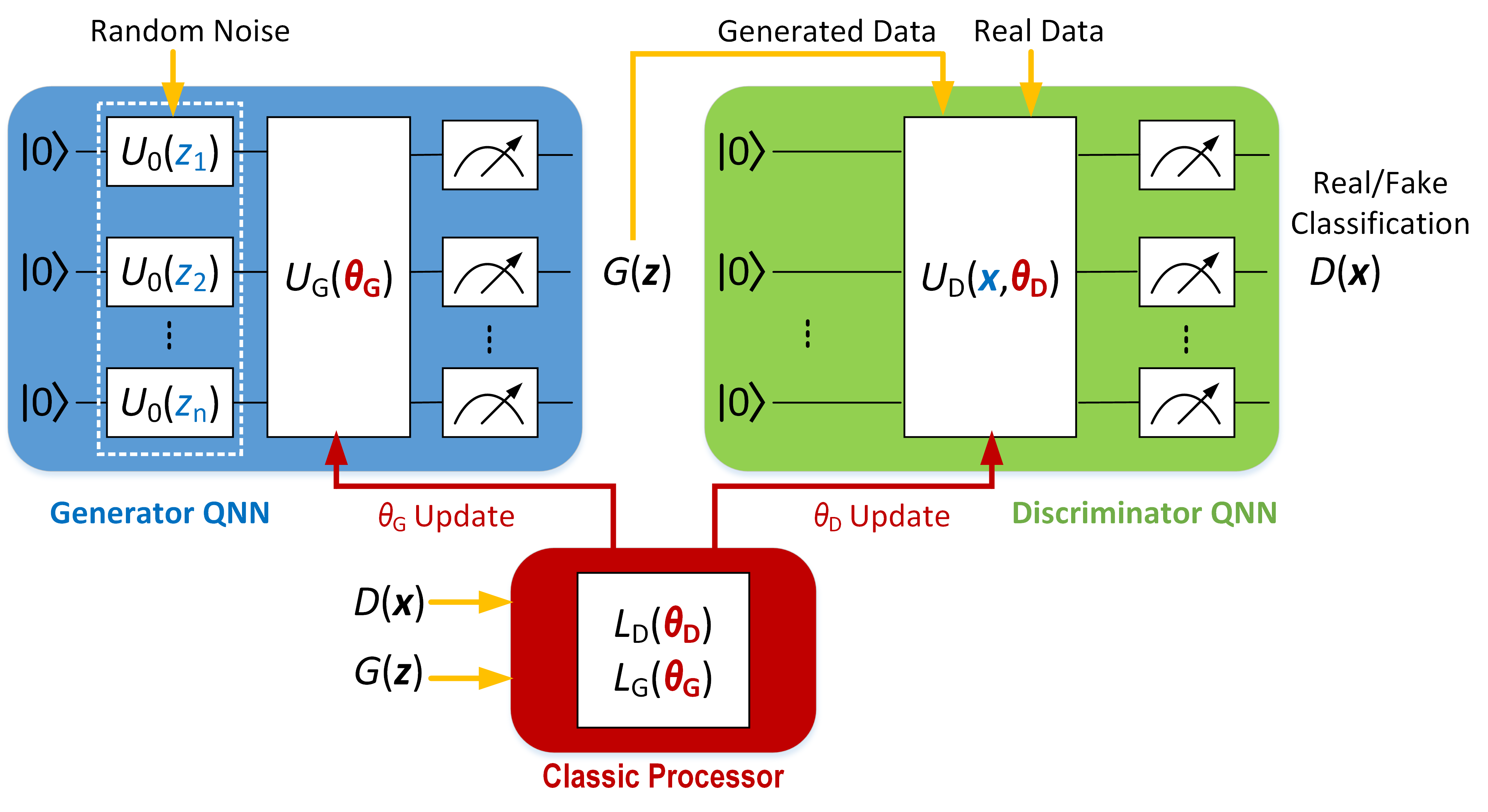}
    \caption{The hybrid quantum-classical GAN architecture. The generator network encodes the random noise into unitary operators $U_0(z_i)$ which combined with the parameterized unitary $U_G({\boldsymbol{\theta_G}})$ and the measurement synthesize the generated data, $G(\boldsymbol{z})$. The generated data along with the real data are fed into the discriminator network which embeds them into the parameterized unitary $U_D(\boldsymbol{x}, \boldsymbol{\theta_D})$ and after the measurement performs the real/fake data classification. The classic processor computes the loss functions $\mathcal{L}_{\text{D}} (\boldsymbol {\theta_D})$ and $\mathcal{L}_{\text{G}} (\boldsymbol {\theta_G})$, and use their optimization results to update the parameters $\boldsymbol{\theta_G}$ and $\boldsymbol{\theta_D}$.} 
    \label{fig:hybrid_GAN}
\end{figure*}

The proposed hybrid quantum-classical GAN architecture is shown in Fig. \ref{fig:hybrid_GAN}. The generator QNN is composed of a quantum encoding circuit and a variational quantum ansatz. The encoding quantum circuit transforms the input classic data into quantum states by using a specific algorithm. The encoding is realized using a unitary quantum operator which can also be regarded as a \textit{feature map} that maps the input data to the Hilbert space of the quantum system. This feature map serves similar to a \textit{kernel} in classic machine learning. The encoding circuit structure can have a significant impact on the generator QNN behavior and should not be overlooked as a trivial operation. The variational quantum ansatz is controlled by the vector parameter $\boldsymbol {\theta_G}$ which is trained on a classic computer. The discriminator QNN is realized as a variational quantum ansatz controlled by the vector parameter $\boldsymbol {\theta_D}$. The generator is trained using a randomized input noise and based on its success in fooling the discriminator. The discriminator is trained using a given dataset and the output of generator until it reaches an acceptable accuracy. 

The generator and discriminator QNNs are trained using a zero-sum game. The loss function for the GAN is defined as

\begin{equation}
\label{loss function}
\begin{aligned}
\mathcal{L}_{\text{GAN}} (\boldsymbol {\theta_D}, \boldsymbol {\theta_G}) = -\E_{x}\left[\log(D(\boldsymbol{x}, \boldsymbol{\theta_D}))\right]  -\\  \E_{z}\left[\log(1-D(G(\boldsymbol{z}, \boldsymbol{\theta_G}), \boldsymbol{\theta_D}))\right]
\end{aligned}
\end{equation}
where $\boldsymbol{x} \sim p_{data}(\boldsymbol{x})$ is the real data, $\boldsymbol{z} \sim p_z(\boldsymbol{z})$ is the noise, $D(\boldsymbol{x}, \boldsymbol{\theta_D})$ denotes the discriminator network, and $G(\boldsymbol{z}, \boldsymbol{\theta_G})$ denotes the generator network. The GAN ideally should solve the minimax game, i.e., the generator tries to minimize the loss function while the discriminator tries to maximize it. In practice, the two networks are trained using separate loss functions \cite{GAN} that should be minimized through iterative optimization
\begin{equation} 
\label{loss_discriminator}
\mathcal{L}_{\text{D}} (\boldsymbol {\theta_D}) = -\E_{x}\left[\log(D(\boldsymbol{x})\right] - \E_{z}\left[\log(1-D(G(\boldsymbol{z})))\right],
\end{equation}
\begin{equation} 
\label{loss_generator}
\mathcal{L}_{\text{G}} (\boldsymbol {\theta_G}) = -  \E_{z}\left[\log(D(G(\boldsymbol{z})))\right],
\end{equation}
where the dependency of the discriminator and generator functions on the training parameters are not shown for simplicity. 
We will discuss the details and practical considerations for the realization and training of the two networks.

\subsection{Structure of Quantum Neural Networks}

A promising approach to realize the QNNs is using VAQs. A VQA is composed of an ansatz which can be realized as a parametric quantum circuit, an input training dataset, and an optimizer which find the optimum values of the ansatz's quantum circuit parameters to minimize a defined cost function. The optimizer usually operates on a classic computer which permits the use of vast knowledge developed on classical machine learning. The ansatz structure, as the core of the VQA, is generally selected based on features of the quantum processor and the specific given task.

The ansatz can be described by a general unitary function $U(\boldsymbol{x}, \boldsymbol{\theta})$, which is conventionally considered as the product of two separate unitary functions $U(\boldsymbol{\theta}) W(\boldsymbol{x})$. The parameter-dependent unitary $U(\boldsymbol{\theta})$ is defined as the product of cascaded unitary functions
\begin{equation}
\label{U}
    U(\boldsymbol{\theta}) = U(\theta_1, ..., \theta_L) = \prod_{k=1}^{L} U_{k}(\theta_k),
\end{equation}
where $U_{k}(\theta_k)$, the unitary function of the layer \textit{k}, is conventionally defined by an exponential function as
\begin{equation}
\label{U_k}
    U_k(\theta_k) = e^{-j H_{k}\theta_{k}}.
\end{equation}
Here, $H_k$ is a Hermitian operator, $\theta_{k}$ is the element $k$ of the training vector $\boldsymbol{\theta}$, and \textit{L} is the number of layers in the ansatz \cite{cerezo2021variational}. The unitary operator (\ref{U_k}) can be interpreted as consecutive parametric rotations with an orientation determined by the Hermitian $H_k$. This is usually selected as one of the Pauli's operators, $H_k \in \{\sigma_x , \sigma_y , \sigma_z \}$, or their linear superposition $\sum_i c_i \sigma_i$. 
The data-dependent unitary function $W(\boldsymbol{x})$ can be similarly modeled as
\begin{equation}
\label{W}
    W(\boldsymbol{x}) = \prod_{i=1}^{N} e^{ - j G_{i}x_{i}},
\end{equation}
where $G_i$ is a constant and \textit{N} is the length of dataset. 
The quantum state prepared by (\ref{U}) is achieved by 
\begin{equation}
\label{PSI}
    \ket{\psi (\boldsymbol{x}, \boldsymbol{\theta})} = U(\boldsymbol{x}, \boldsymbol{\theta}) \ket{\psi_0},
\end{equation}
where $\ket{\psi_0}$ is an initial quantum state. This quantum state can be used to define a quantum model.

Classical optimization problems in the VQAs are generally in the NP-hard complexity class \cite{bittel2021}, and, therefore, should be solved using specifically developed methods, e.g., adapted stochastic gradient descent (SGD) \cite{sweke2020} or meta-learning \cite{meta_learning, verdon2019learning}. 
Major limitations of the VQAs are related to their trainability, efficiency, and accuracy. Trainability of the VQAs is limited by the Barren plateaus phenomenon (Section II-D), where gradient of the cost function can become exponentially vanishing for a range of training parameters \cite{mcclean2018barren}. Efficiency of the VQAs is evaluated by the number of measurements required to estimate expectation values in the cost function. Generally, the efficiency is dependent on the computational task of the VQA. Accuracy of the VQAs is mostly impacted by hardware noise which is a notorious feature of the NISQ processors. Noise slows down the training process, degrades accuracy of the algorithm by impacting final value of the cost function, and can lead to the noise-induced barren plateaus \cite{wang2021noise}. These practical challenges should be considered for efficient architecture development and training of the VQAs. 

Conventionally the VQAs are realized using parameterized quantum circuits where the parameters specify the rotation angles for the quantum gates. These angles are trained to minimize a specified loss function defined for a given computational task. The loss function is dependent on the system inputs $x_k$, circuit observables denoted by $O_k$, and trainable parameterized unitary circuit $U(\boldsymbol{\theta})$
\begin{equation}
\label{C_theta}
    \mathcal{L}(\boldsymbol{\theta}) = f({x_k}, {O_k}, U(\boldsymbol{\theta})).
\end{equation}
The ansatz parameters are optimized using a classic optimizer to minimize the cost function
\begin{equation}
\label{theta_opt}
    \boldsymbol{\theta^*} = \underset{\boldsymbol{\theta}} {\arg\min}\, \mathcal{L}(\boldsymbol{\theta}).
\end{equation}

The choice of ansatz architecture is an important consideration for QNN operation. There are some generic quantum circuit architectures which have performed well on a variety of computational problems and conventionally are used in other applications \cite{cerezo2021variational, zhao2021review}. These ansatzes have a fixed architecture and their trainable parameters are usually rotation angles. These algorithms are relatively easy to implement and are hardware efficient, as they enable control of the circuit complexity. We implement the generator QNN using this approach. On the other hand, the quantum circuit architecture can also be modified to find optimal architecture for the given task. We use this quantum neural architecture search approach to realize the discriminator QNN. 

The number of layers is an important design parameter of a QNN. Unfortunately, there is no well-defined rule for finding the optimum number of layers. In a fault-tolerant quantum computer, a deep QNN comprising large number of layers, as allowed by the processor computational power, can provide higher accuracy. However, in the NISQ computers performance of the quantum algorithm can degrade by the decoherence of qubits, especially if the QNN includes many layers. Therefore, in this work we use a \textit{modular design approach} to control \textit{depth} of QNNs. This can mitigate the impacts of noise and barren plateaus, and improve trainability of the QNNs.

\subsection{Gradient Evaluation}

The cost function in VQAs is derived using hybrid quantum-classical processing which should be minimized through a classical optimization approach. It is useful to derive exact gradients of quantum circuits with respect to the gate parameters. Using the \textit{parameter shift rule}, gradients of a quantum circuit can be estimated using the same architecture that realizes the original circuit
 \begin{equation} \label{param_shift}
\nabla_{\theta}f(x, \theta) = k [f(x, \theta + \Delta \theta) - f(x, \theta - \Delta \theta) ],
\end{equation}
where $\Delta \theta$ denotes parameter shift and $k$ is a constant \cite{schuld2019evaluating}. This reduces the computational resources required to evaluate gradient of quantum circuits.

A quantum rotation gate with a single-variable rotation $\theta$ can be written in the form $G(\theta) = e^{-j h \theta}$, where $h$ is a parameter representing the behavior of the gate. It is straightforward to show that the first-order derivative of this function can be derived as 
\begin{equation} \label{deriv_gate}
    \frac{\partial G(\theta)}{\partial \theta} = k \left(G \left( \theta + \frac{\pi}{2} \right) - \left( \theta - \frac{\pi}{2} \right) \right).
\end{equation}
This leads to the conclusion that gradient of a quantum circuit composed of rotation gates can be derived from the same quantum circuit evaluated at the shifted angles $\theta \pm \frac{\pi}{2}$. This result can be extended to a quantum circuit which its Hermitian operator has two distinct eigenvalues \cite{mitarai2018quantum, schuld2019evaluating}. For a quantum circuit with a measurement operator $\hat{O}$ and the quantum state $ \ket{\psi (\boldsymbol{\theta})}$, as defined in (\ref{PSI}), the first-order derivative of the output quantum state $\braket{\psi(\boldsymbol{\theta})| \hat{O} | \psi(\boldsymbol{\theta})}$ with respect to the element $\theta_i$ of the vector $\boldsymbol{\theta} = \{\theta_1, ..., \theta_i, ..., \theta_n\}$ can be derived as

\begin{equation} \label{full_grad}
\begin{split}
    \frac{\partial \braket{\psi(\boldsymbol{\theta})| \hat{O} | \psi(\boldsymbol{\theta})}}{\partial \theta_i} = \frac{1}{2} \left(\braket{\psi(\boldsymbol{\theta_i^{+}})| \hat{O} | \psi(\boldsymbol{\theta_i^{+}})} - \right. \\* \left. \braket{\psi(\boldsymbol{\theta_i^{-}})| \hat{O} | \psi(\boldsymbol{\theta_i^{-}})}\right),
\end{split}
\end{equation}
where 
\begin{equation}
    \label{theta_shifted}
    \boldsymbol{\theta_i^{\pm}} = \{\theta_1, ..., \theta_i \pm \frac{\pi}{2}, ..., \theta_n\}.
\end{equation}

All of the angular rotation gates used in this paper feature Hermitian operators with two distinct eigenvalues. Therefore, the gradient calculation approach can be applied. It is noted that this method requires the output to be calculated twice for each trainable parameter. On the other hand, it doesn't need any ancila qubit as in other methods \cite{izmaylov2021analytic}. In the hybrid GAN algorithm, there is a large number of trainable parameters in the generator and discriminator QNNs and, as a result, the use of parameter shift rule can obviate the need for using many ancila qubits.

\subsection{Barren Plateaus}
Barren plateaus issue is one of the challenges in training process of hybrid quantum-classical algorithms where a parameterized quantum circuit is optimized by a classical optimization loop. This results in exponentially vanishing gradients over a wide landscape of parameters in QNNs \cite{mcclean2018barren, cerezo2021cost, Grant2019initialization, verdon2019learning, hadfield2019quantum}. Mechanisms of the barren plateaus are not completely understood and its behaviors are still under investigation. In \cite{mcclean2018barren}, it is shown that deep quantum circuits with random initialization are especially prone to the barren plateaus. Furthermore, the number of qubits plays an important role and for sufficiently large number of qubits there is a high chance that the QNN training end up to a barren plateau. 

A number of strategies have been proposed to mitigate the effects of barren plateaus in the training of QNNs. In \cite{Grant2019initialization}, it is proposed to randomly initialize the parameters of the first circuit block $U_k({\theta}_{k,1})$ and add a subsequent block whose parameters are set such that the unitary evaluates to $U_k({\theta}_{k,2}) = U_k({\theta}_{k,1})^\dagger$. This means that each circuit block and therefore the whole circuit will initially evaluate to the identity, i.e., $U_k({\theta}_{k,1}) U_k({\theta}_{k,1})^\dagger = 1$. This technique can be effective in reducing the effects of barren plateaus, but it can alter the quantum circuit structure depending on the initial choice of the ansatz, which could degrade the generalization of the quantum classifier. Other approaches including the use of structured initial guesses and the segment-by-segment pretraining can also mitigate this issue \cite{mcclean2018barren}. 

In the hybrid quantum-classical GAN proposed in this paper, there are at least two features which reduce the barren plateaus effect. First, the modular structure of the QNNs with a primarily advantage of controlling their depth to improve fidelity of the quantum circuits realized using noisy qubits, can also reduce the gradient complexity and the barren plateaus. Second, the modular design approach used to find optimal QNN structures leads to a trainable encoding scheme (Section III-B) which can help to identify the network architectures with the barren plateaus problem and modify them to resolve the issue.

\subsection{Discriminator Network} \label{section:classifier}

The discriminator QNN of the hybrid quantum-classical GAN architecture is composed of a trainable encoding circuit and separate layers of a parameterized circuit ansatz. Details of the encoding circuit are discussed in Section III. 
The discriminator network is used to realize a quantum variational classifier. The expectation value of a single qubit in the discriminator QNN is used to compute the neural network output. The output of the discriminator network is a real number between $0$ and $1$ indicating the probability of an input being a real data sample. The Z-basis expectation value measured at the output of the discriminator network is given by $\braket{\psi | \sigma_z | \psi}$, where $\sigma_z$ is the Pauli-Z matrix. Assuming $\ket{\psi} = \alpha \ket{0} + \beta \ket{1}$, where $\alpha$ and $\beta$ are complex numbers satisfying $|\alpha|^2 + |\beta|^2 =1$, it can be shown that
\begin{equation}
    \label{Z_meas}
    \braket{\psi | \sigma_z | \psi} = |\alpha|^2 - |\beta|^2.
\end{equation}
This quantity changes from $-1$ for $\ket{\psi} =  \ket{1}$ to $1$ for $\ket{\psi} =  \ket{0}$. We can scale it to achieve a number between $0$ and $1$ at output of the discriminator network
\begin{equation}\label{normalise_discriminator}
D = \frac{1}{2} \left( 1 +  \braket{\psi|\sigma_z|\psi} \right).
\end{equation}

The loss function for the discriminator network is defined in (\ref{loss_discriminator}). The loss can be minimized using the minibatch stochastic gradient descent (SGD). This requires the gradient of the loss function $\mathcal{L}_{\text{D}} (\boldsymbol {\theta_D})$ to be evaluated with respect to the vector parameter $\boldsymbol {\theta_D}$. Using (\ref{loss_discriminator}), the derivative of the loss function with respect to the component $\theta_D^i$ of the vector can be derived as
\begin{equation}
    \label{LD_derivative}
    \frac{\partial \mathcal{L}_{\text{D}} (\boldsymbol {\theta_D})}{\partial \theta_D^i} = \frac{\partial \mathcal{L}_{\text{D}} }{\partial D} \frac{\partial D}{\partial \braket{\psi|\sigma_z|\psi}} 
    \frac{\partial \braket{\psi|\sigma_z|\psi}}{\partial \theta_D^i}.
\end{equation}
There are two terms in the discriminator loss function (\ref{loss_discriminator}) which we denote them as $\mathcal{L}_{\text{D1}}$ and $\mathcal{L}_{\text{D2}}$. 
The first derivative term of (\ref{LD_derivative}) can be derived using (\ref{loss_discriminator}) as
\begin{equation}
    \label{term1}
    \frac{\partial \mathcal{L}_{\text{D1}} }{\partial D} = - \frac{1}{D(\boldsymbol{x}, \boldsymbol{\theta_D})}
\end{equation}
\begin{equation}
    \label{term1_2}
    \frac{\partial \mathcal{L}_{\text{D2}} }{\partial D} =  \frac{1}{1 - D(G(\boldsymbol{z}), \boldsymbol{\theta_D})}.
\end{equation}
The second derivative term using (\ref{normalise_discriminator}) is given by
\begin{equation}
    \label{term2}
    \frac{\partial D}{\partial \braket{\psi|\sigma_z|\psi}} = \frac{1}{2}.
\end{equation}
The third derivative term can be calculated using the parameter shift rule, (\ref{full_grad}) and (\ref{theta_shifted}), as
\begin{equation}
    \label{term3}
      \frac{\partial \braket{\psi|\sigma_z|\psi}}{\partial \theta_D^i} = \\ \frac{1}{2}  \left( \braket{\psi(\boldsymbol{\theta_{D}}^{i+})|\sigma_z|\psi(\boldsymbol{\theta_{D}}^{i+})}  -  \braket{\psi(\boldsymbol{\theta_{D}}^{i-})|\sigma_z|\psi(\boldsymbol{\theta_{D}}^{i-})}  \right),
\end{equation}
where 
\begin{equation}
    \label{theta_D_shifted}
    \boldsymbol{\theta_D^{i\pm}} = \{\theta_D^1, ..., \theta_D^i \pm \frac{\pi}{2}, ..., \theta_D^n\}.
\end{equation}
Comparing (\ref{term3}) with (\ref{normalise_discriminator}) indicates that the third derivative term can be related to the discriminator output as the difference between its value evaluated at $\boldsymbol{\theta_D^{i+}}$ and $\boldsymbol{\theta_D^{i-}}$. Therefore, using (\ref{LD_derivative})--(\ref{theta_D_shifted}), the derivative of the two terms of the loss function can be derived as
\begin{equation}
    \label{LD_derivative_v2}
    \frac{\partial \mathcal{L}_{\text{D1}} (\boldsymbol {\theta_D})}{\partial \theta_D^i} = - \frac{1}{2} \frac{D(\boldsymbol{x}, \boldsymbol{\theta_D^{i+}}) - D(\boldsymbol{x}, \boldsymbol{\theta_D^{i-}})}{D(\boldsymbol{x},\boldsymbol{ \theta_D})}
\end{equation}
\begin{equation}
    \label{LD_derivative_v2_2}
    \frac{\partial \mathcal{L}_{\text{D2}} (\boldsymbol {\theta_D})}{\partial \theta_D^i} =  \frac{1}{2} \frac{D(G(\boldsymbol{z}), \boldsymbol{\theta_D^{i+}}) - D(G(\boldsymbol{z}), \boldsymbol{\theta_D^{i-}})}{1 - D(G(\boldsymbol{z}), \boldsymbol{\theta_D})}
\end{equation}
This process can be performed for all elements of the vector parameter $\boldsymbol {\theta_D}$ to derive gradient of the discriminator loss function as
\begin{equation}
    \label{gradient_disc}
    \boldsymbol{\nabla} \mathcal{L}_{\text{D}} = \sum_{i=1}^{n}{\frac{\partial \mathcal{L}_{\text{D}} (\boldsymbol {\theta_D})}{\partial \theta_D^i} \hat{u}_D^i},
\end{equation}
where $\hat{u}_D^i$ is the unit vector in the direction of parameter $\theta_D^i$.

\subsection{Generator Network}
The generator QNN of the hybrid quantum-classical GAN architecture is composed of a quantum encoding circuit and a parameterized circuit ansatz. A fixed encoding scheme is used in the generator network unlike the discriminator network which included a trainable encoding circuit. Details of this approach are presented in Section III. 
Similar to the discriminator network, the Pauli-Z operator is used as the basis for measurement of the expectation values of output qubits in the generator network, i.e., $\braket{\psi | \sigma_z | \psi}$. The expectation values can be normalized using (\ref{normalise_discriminator}) to constraint output of the generator network between $0$ and $1$. 

The loss function for the generator network is defined in (\ref{loss_generator}). The network is trained using the minibatch SGD and the loss associated with each batch is derived by averaging across all data samples in that batch. Using (\ref{loss_generator}), the derivative of the loss function  $\mathcal{L}_{\text{G}} (\boldsymbol {\theta_G})$ with respect to the component $\theta_G^i$ of the vector $\boldsymbol {\theta_G}$ can be derived as
\begin{equation}
    \label{LG_derivative}
    \frac{\partial \mathcal{L}_{\text{G}} (\boldsymbol {\theta_G})}{\partial \theta_G^i} = \frac{\partial \mathcal{L}_{\text{G}} }{\partial D} \frac{\partial D}{\partial G} 
    \frac{\partial G}{\partial \theta_G^i}.
\end{equation}
The first term can be directly calculated as
\begin{equation}
    \label{term1G}
    \frac{\partial \mathcal{L}_{\text{G}} }{\partial D} = - \frac{1}{D(G(\boldsymbol{z}), \boldsymbol{\theta_D})}.
\end{equation}
For the second term, we note that the output of the discriminator network is related to the generator network function as $D = D(G(\boldsymbol{z}, \boldsymbol{\theta_G}), \boldsymbol{\theta_D})$, where $\boldsymbol{\theta_D}$ is treated as a constant when the generator network is evaluated. The general form of the parameter shift rule, defined by (\ref{param_shift}), can be applied provided that 
\begin{equation}
    \label{term2G}
    \frac{\partial D}{\partial G} = k_G \left( D(G + \Delta G) - D(G - \Delta G) \right),
\end{equation}
where $k_G$ and $\Delta G$ are two properly selected parameters. $G(\boldsymbol{z}, \boldsymbol{\theta_G})$ is controlled by the vector parameter $\boldsymbol{\theta_G}$ and, therefore, $k_G$ and $\Delta G$ should also be generally dependent on it. These parameters cannot be explicitly determined without a complete information about the $G(\boldsymbol{z}, \boldsymbol{\theta_G})$ function. A solution is to initialize their value based on (\ref{full_grad}) and (\ref{theta_shifted}), i.e., $k_G^{(1)} = \frac{1}{2}$ and $\Delta G^{(1)} = \frac{\pi}{2}$, and then optimize them during the overall SGD process. 

Furthermore, third term in (\ref{LG_derivative}) can be derived using the parameter shift rule, (\ref{full_grad}) and (\ref{theta_shifted}), as
\begin{equation}
    \label{term3G}
    \frac{\partial G}{\partial \theta_G^i} = \frac{1}{2} \left( G(\boldsymbol{z}, \boldsymbol{\theta_G^{i+}}) - G(\boldsymbol{z}, \boldsymbol{\theta_{G}^{i-}}) \right),
\end{equation}
where 
\begin{equation}
    \label{theta_G_shifted}
    \boldsymbol{\theta_G^{i\pm}} = \{\theta_G^1, ..., \theta_G^i \pm \frac{\pi}{2}, ..., \theta_G^n\}.
\end{equation}
The gradient of the generator loss function can be derived by running this process for all elements of the vector parameter $\boldsymbol {\theta_G}$ as
\begin{equation}
    \label{gradient_gen}
    \boldsymbol{\nabla} \mathcal{L}_{\text{G}} = \sum_{i=1}^{n}{\frac{\partial \mathcal{L}_{\text{G}} (\boldsymbol {\theta_G})}{\partial \theta_G^i} \hat{u}_G^i},
\end{equation}
where $\hat{u}_G^i$ is the unit vector in the direction of parameter $\theta_G^i$.

\section{Data Encoding}

\subsection{Fundamentals}
Data representation is a critical aspect for performance of quantum machine learning. Classical data must first be transformed into quantum data before it can be processed by quantum or hybrid quantum-classical algorithms. The data encoding can be realized using different schemes, including basis encoding, angle encoding, and amplitude encoding \cite{robust_encoding, schuld2019quantum, cortese2018loading}. In the NISQ computers, trainability and robustness of an algorithm are dependent on the encoding scheme. 

The encoding is conventionally realized using certain quantum circuits with a fixed structure (Section III-B). In quantum machine learning, this approach prepares the data for training of the QNN independent of the trainable parameters of the network. We use this approach for the generator network. However, in the discriminator network, we use a \textit{trainable encoding} scheme which improves the ultimate accuracy (Section III-C).

\subsection{Fixed Encoding Schemes}

The data encoding entails mapping the classic data $\boldsymbol{x}$ to a high-dimensional quantum Hilbert space using a quantum feature map $\boldsymbol{x} \rightarrow \ket{\psi(\boldsymbol{x})}$ \cite{schuld2019quantum, cortese2018loading, quantum_feature}. We evaluate different encoding approaches for the hybrid quantum-classical GAN application on the NISQ computers\footnote{Three fundamental encoding methods are investigated in this paper.}.

\subsubsection{Basis Encoding}

In the basis encoding, the classical data is converted to the binary representation  $\boldsymbol{x} \in \{0,1\}^{\otimes N}$ and then is encoded as the basis of a quantum state 
\begin{equation}
\label{basis_encoding}
    \ket{\boldsymbol{x}} = \frac{1}{\sqrt{N}}\sum_{i=1}^{N} \ket{x_i}.
\end{equation}
The number of qubits is the same as the number of bits in the classical data representation, $Q = \log_2{N}$. The basis encoding transforms the data samples into quantum space where qubits can store multiple data samples using the quantum superposition. However, the realization of basis encoding requires additional auxiliary qubits \cite{cortese2018loading}. This is particularly undesirable for the NISQ computers as the excessive qubits can degrade the fidelity of quantum algorithms.

\subsubsection{Angle Encoding}

In the angle encoding, the classic data $\boldsymbol{x} = [x_1, ..., x_N]^T$ is embedded into the angle of qubits as
\begin{equation} \label{angle_encoding}
   \ket{\boldsymbol{x}} = \bigotimes_{i = 1}^{N} \cos(x_i)\ket{0} + \sin(x_i)\ket{1}.
\end{equation}
The realization of this approach requires the same number of qubits as the number of data samples, $Q = N$ qubits. It is therefore less qubit efficient compared to the basis and amplitude encoding methods. On the other hand, the angle encoding can be implemented using a \textit{constant depth quantum circuit}, e.g., single-qubit rotation gates, e.g., RX, RY, RZ, and is therefore amenable to the NISQ computers.

In the angle encoding, unlike the basis and amplitude encoding methods, the data features are consecutively encoded into the qubits. This increases the data loading time by a factor of $N$. It can be resolved by using $N$ parallel quantum gates. The basic angle encoding defined by (\ref{angle_encoding}) can be modified to the dense angle encoding \cite{robust_encoding} where two data features are encoded per qubit 
\begin{equation} \label{dense_angle_encoding}
   \ket{\boldsymbol{x}} = \bigotimes_{i = 1}^{N/2} \cos(\pi x_{2i-1})\ket{0} + e^{j2\pi x_{2i}}\sin(\pi x_{2i-1})\ket{1}.
\end{equation}
This methods needs $Q = \frac{1}{2}N$ qubits and improves the data loading speed. However, a more complicated quantum circuit is required to implement the encoding. 

A subtle important point is that the angle encoding performs \textit{nonlinear operations} on the data through the sine and cosine functions. This feature map can provide powerful \textit{quantum advantage} in quantum machine learning, e.g., separate the data such that simplify the QNN circuit architecture and improve its performance.

\subsubsection{Amplitude Encoding}

The amplitude encoding embeds the classic data into the amplitudes of a quantum state
\begin{equation} \label{ampl_enc}
    \ket{\boldsymbol{x}} = \frac{1}{||\boldsymbol{x}||} \sum_{i =1}^Nx_i\ket i ,
\end{equation}
where $||\boldsymbol{x}|| = \sqrt{\sum_i \left|x_i\right|^2} $ denotes the Euclidean norm of $\boldsymbol{x}$. The number of data features $N$ is assumed to be a power of two, $N = 2^n$, and if necessary, this can be achieved by zero padding. This encoding method requires $Q = \log_2 {N}$ qubits. Quantum circuits used to realize the amplitude encoding do not have a constant depth \cite{lloyd2020quantum}. Therefore, the amplitude encoding suffers the circuit implementation complexity and a higher computational cost compared to the angle encoding. These considerations make the amplitude encoding less favorable for the NISQ computers.

\subsubsection{Encoding Circuit Implementation}

For each of the presented data encoding methods, there are many quantum circuits which can be used to realize that encoding. The circuits can feature different number of qubits, auxiliary qubits, quantum gates, and the circuit depth. The problem of finding the optimal encoding scheme for a given computational task still has not been explored in the literature. The encoding circuit can be selected based on some criteria including the number of required qubits and auxiliary qubits, computation cost, trainability and robustness to noise other imperfections in the quantum computer hardware. 

The generator network of the GAN should provide a broad diversity of output synthetic quantum data to be fed to the discriminator network. There is no certain desired output data distribution to be used as the basis for optimization of the input encoding circuit. Therefore, we select a fixed encoding quantum circuit followed by a parameterized circuit ansatz as the generator network (Section IV-A). The angle encoding is realized using multiple parallel rotation gates.

\subsection{Trainable Encoding Schemes}

The fixed encoding schemes treat the encoding task merely as a classical-to-quantum data transformation for subsequent processing by quantum circuits. The data encoding is performed completely independent of the trainable parameterized quantum circuits. We can envision a different avenue by directly embedding the classical data into the quantum circuit as their control parameters. This allows the data to be encoded using hardware-efficient quantum circuits which are conjectured to be hard to simulate classically for data encoding \cite{quantum_feature, lloyd2020quantum, rotteler2010quantum}. Furthermore, the nonlinear operations which can be realized using encoding circuits, e.g., the angle encoding discussed in Section III-B, can be included in different layers of the quantum circuit. The interleaving of the rotation gates conditioned by the input data with the variational rotation gates creates a trainable encoding circuit which can be trained based on the given input data and expected output. It is shown that this approach can reduce complexity of the required parameterized quantum circuit or, in some cases, the trainable encoding circuit can accomplish the certain task without using any additional circuit ansatz \cite{lloyd2020quantum}. The trainable encoding schemes can provide unique properties which call for future research.

In the developed hybrid quantum-classical GAN in this paper, we have implemented the discriminator network using the trainable encoding approach. Two trainable encoding circuits and an ansatz circuit are used as the possible building blocks of the discriminator network. The discriminator network architecture is selected by evaluating multiple possible architectures achieved through different combinations of the encoding and ansatz circuits (Section IV-B).

\section{QNN Architectures}

\subsection{Generator QNN}
\begin{figure}[!t]
    \centering
    \includegraphics[width = \columnwidth]{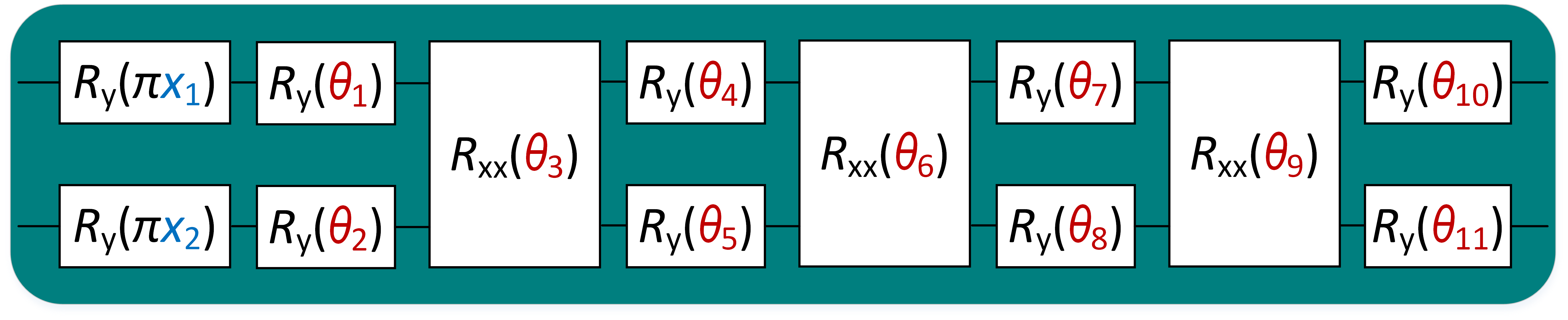}
    \caption{Architecture of the generator QNN. The data encoding is realized using the RY gates with inputs $x_i$ and the variational ansatz circuit includes the trainable parameters $\theta_i$.}
    \label{fig:gen_qnn}
\end{figure}

The generator QNN architecture is shown in Fig. \ref{fig:gen_qnn} which comprises a fixed encoding quantum circuit followed by a parameterized circuit ansatz. The angle encoding is realized using parallel RY gates with scaled input data features $R_y(\pi x_i)$ to control the range of rotation angle. The RY gate, defined as $R_y(\theta) = \exp{(-j\frac{\theta}{2} \sigma_y)}$, performs the rotation about the Y axis. For example, $R_y(\pi x_i)$ transforms the initial qubit state $\ket{0}$ to $\ket{\psi_i} = \cos(\frac{\pi}{2}x_i) \ket{0} + \sin{(\frac{\pi}{2}x_i)} \ket{1}$. 

The variational ansatz circuit is realized using parameterized RY and entangling RXX gates. The RXX gate is defined as $R_{xx}(\theta) = \exp{(-j\frac{\theta}{2} \sigma_x \otimes \sigma_x)}$, with the maximum entanglement at $\theta = \frac{\pi}{2}$ and the identity matrix (no entanglement) at $\theta = 0$. The generator QNN includes a total of 11 trainable parameters $\theta_i$.

\subsection{Discriminator QNN}
\begin{figure}[!t]
    \centering
    \includegraphics[width = 1.0\columnwidth]{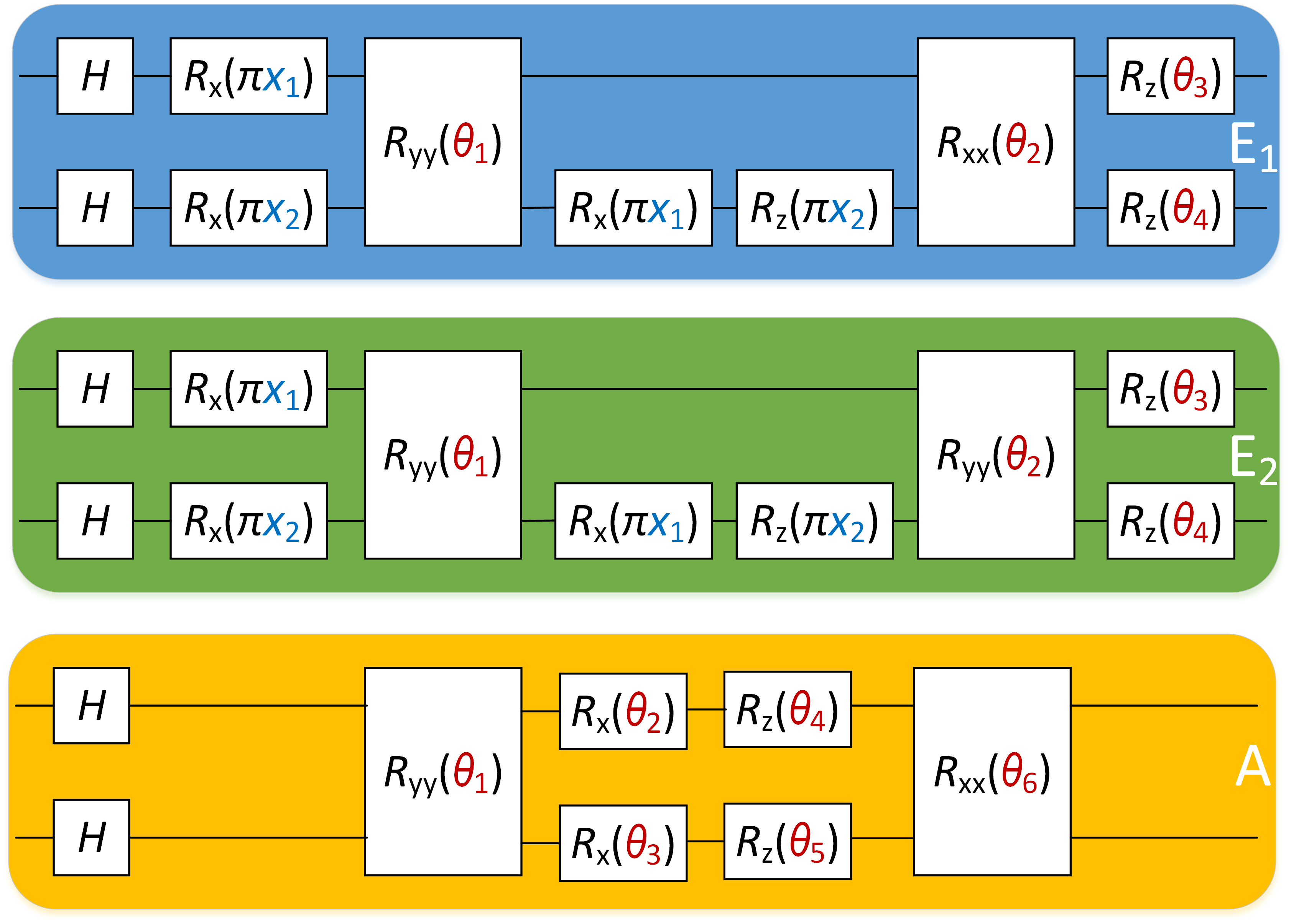}
    \caption{Trainable encoding circuits ($\rm E_1$ and $\rm E_2$) and parameterized ansatz circuit ($\rm A$) for the realization of discriminator network. The data features embedded in the circuits are shown by $x_i$ and the trainable parameters of the quantum gates are denoted by $\theta_i$.}
    \label{fig:quantum_circuits}
\end{figure}

The discriminator QNN is realized using a combination of trainable encoding circuits and a parameterized ansatz circuit. We consider two trainable encoding circuits and the ansatz circuit shown in Fig. \ref{fig:quantum_circuits} as the possible building blocks of the discriminator network. 

In the first encoding circuit, $\rm E_1$ in Fig. \ref{fig:quantum_circuits}, the first layer comprises the Hadamard gates which transform the input qubits state to superposition of the basis states with equal probabilities. The second layer includes the RX gates with $\pi$-scaled input data features $x_1$ and $x_2$. The RX gate, given by $R_x(\theta) = \exp{(-j\frac{\theta}{2} \sigma_x)}$, performs rotation about the X axis. The third layer is a two-qubit RYY gate which creates entanglement between its input qubits. The gate is defined as $R_{yy}(\theta) = \exp{(-j\frac{\theta}{2} \sigma_y \otimes \sigma_y)}$, providing the maximum entanglement at $\theta = \frac{\pi}{2}$ and the identity matrix at $\theta = 0$. The next layers are consecutive RX and RZ gates with scaled input data features $x_1$ and $x_2$ operating on one of the qubits from the previous gate.  The RZ gate, given by $R_z(\theta) = \exp{(-j\frac{\theta}{2} \sigma_z)}$, applies rotation about the Z axis. The next layer is a parameterized RXX entangling gate and the last layer is realized using the RZ gates.

The second encoding circuit, $\rm E_2$ in Fig. \ref{fig:quantum_circuits}, has a same architecture as the first encoding circuit except for the RXX gate being replaced with an RYY gate. This modifies properties of the entanglement generated between the two qubits. 

The ansatz circuit, $\rm A$ in Fig. \ref{fig:quantum_circuits}, is realized using five layers including six trainable quantum gates. The circuit architecture is selected such that include the fixed Hadamard gates, trainable single-qubit rotation gates RX and RZ, and trainable two-qubit entangling gates RXX and RYY.

\begin{figure}[!t]
    \centering
    \includegraphics[width = 0.9\columnwidth]{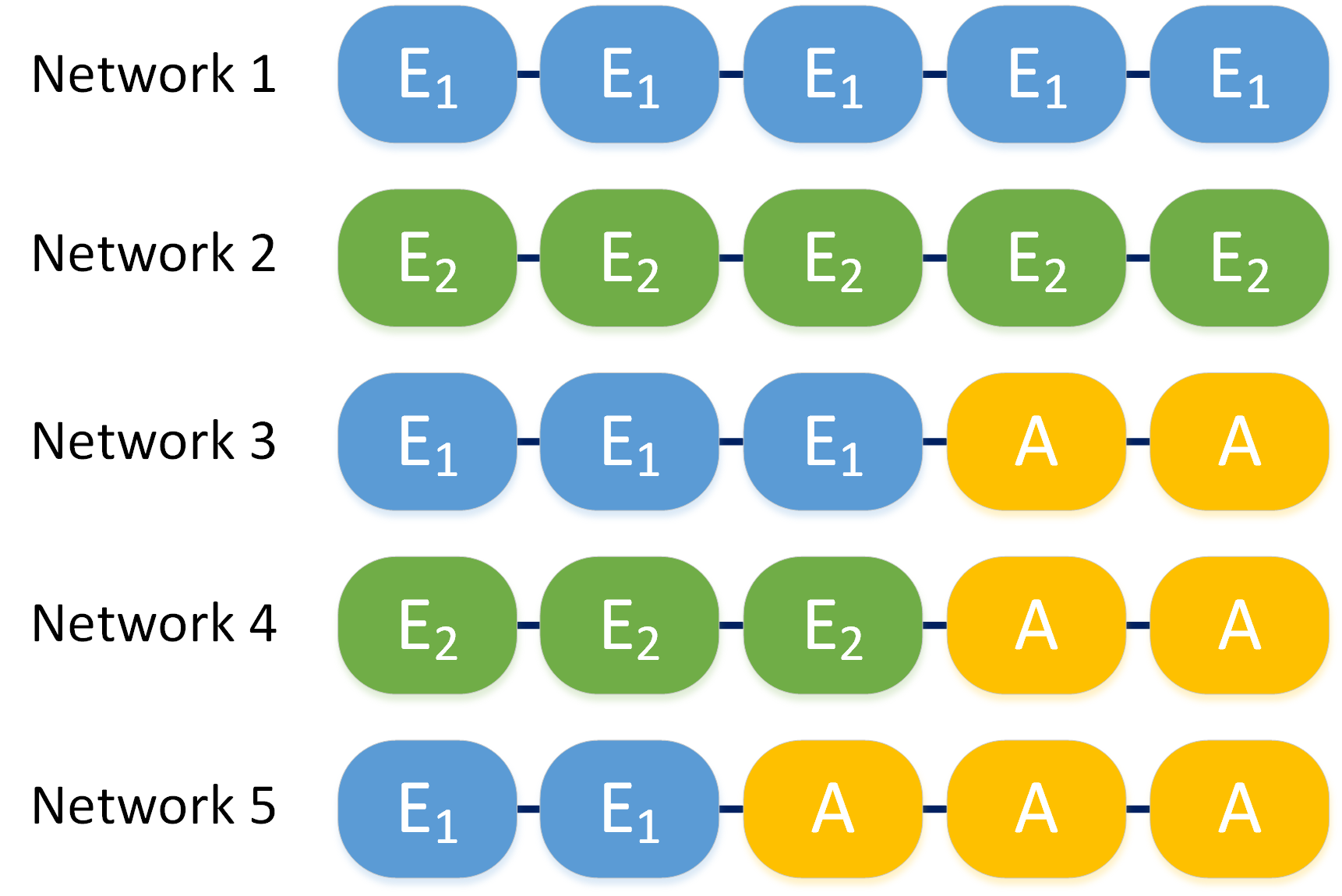}
    \caption{Five quantum network architectures considered as candidates for the realization of discriminator QNN. Details of the encoding and ansatz quantum circuits $\rm E_1$, $\rm E_2$, and $\rm A$ are shown in Fig. \ref{fig:quantum_circuits}.}
    \label{fig:qnn_options}
\end{figure}

The discriminator QNN has been trained using the two-moon dataset which is well-suited to the evaluation of quantum machine learning algorithms on the NISQ computers \cite{schuld2020circuit, lloyd2020quantum}. This is a two-feature dataset with a crescent shape decision boundary. The dataset samples are normalized to the range 0 to 1. We have explored an extensive number of QNN architectures realized using different combinations of the encoding and ansatz circuits of Fig. \ref{fig:quantum_circuits}. The networks are evaluated based on their training time and accuracy. Five representative networks considered as candidates for the discriminator QNN are shown in Fig. \ref{fig:qnn_options}. This \textit{modular design approach} enables control on trade-offs between accuracy of circuit complexity of the QNNs which is particularly important for the NISQ processors.

\begin{figure}[!t]
    \centering
    \includegraphics[width = 0.9\columnwidth]{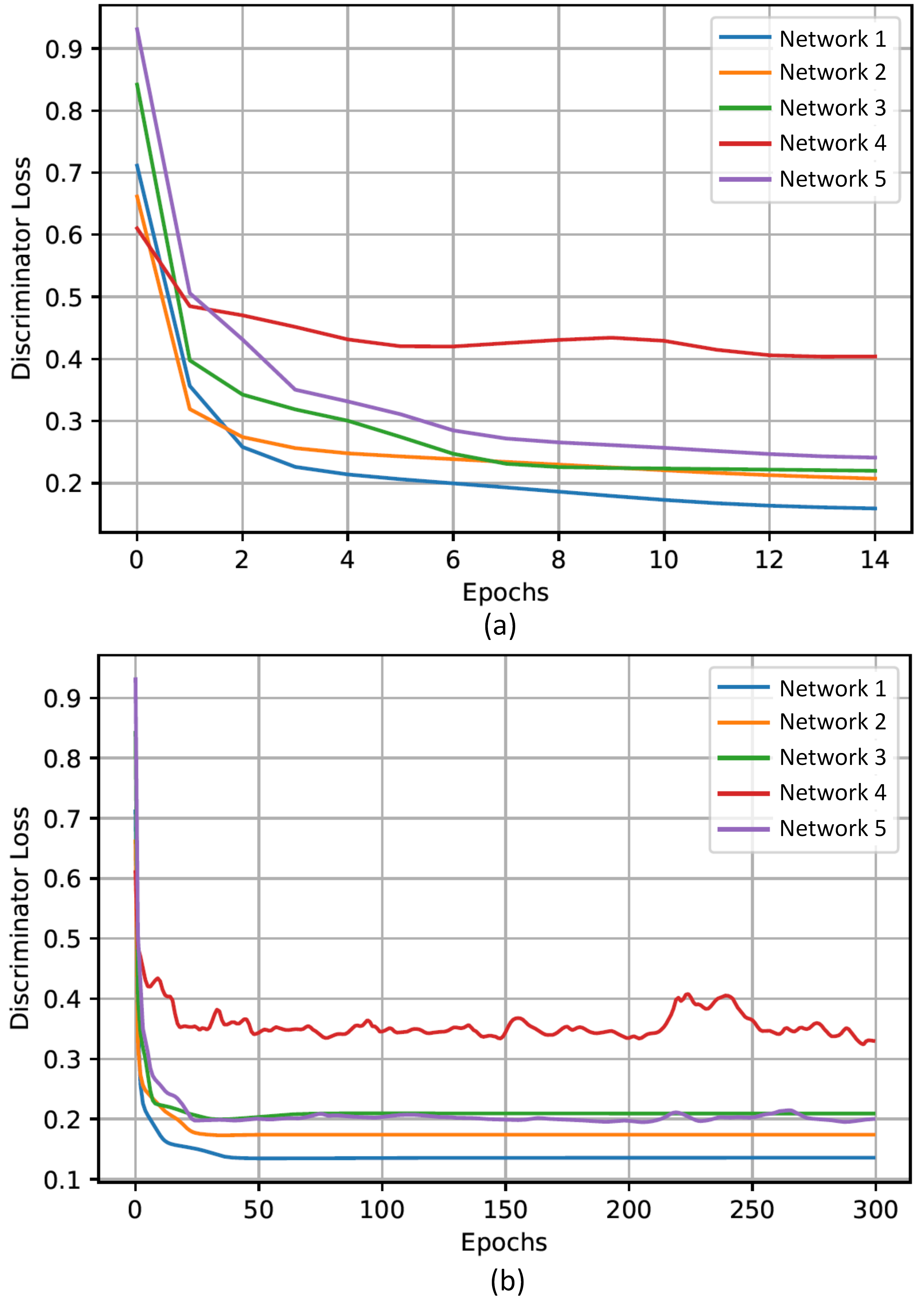}
    \caption{The discriminator network training loss for different architectures: (a) up to 14 epochs, (b) up to 300 epochs.}
    \label{fig:disc_qnn_loss}
\end{figure}

The discriminator loss during training using the two-moon dataset for different network candidates is shown in Fig. \ref{fig:disc_qnn_loss}. The networks 1 and 2, which are comprised of only trainable encoding circuits, achieve the lowest loss levels (0.14 and 0.18, respectively). This validates our intuition about the power of trainable encoding schemes discussed in Section III-C. Furthermore, it is noted that the network 1 converges to the final loss level faster than the network 2. The two networks differ in one of their entangling layers (Fig. \ref{fig:quantum_circuits}), but this leads to different behaviors in their training dynamics. This conclusion highlights the importance of using proper quantum gates to create the entanglement. 

The networks 3, 4, and 5 which comprise the ansatz circuit as well as one of the trainable encoding circuits achieve higher loss levels, in spite of using more trainable quantum gates. The network 4 reaches a loss of 0.35 after only about 20 training epochs and fluctuates around this loss level upon further training. This can be an indicator of the barren plateaus which prevent efficient training of the network. It is concluded that adding more trainable gates to the network architecture not only necessarily improves its accuracy, but also can degrade the accuracy. 

\begin{figure}[!t]
    \centering
    \includegraphics[width = 0.9\columnwidth]{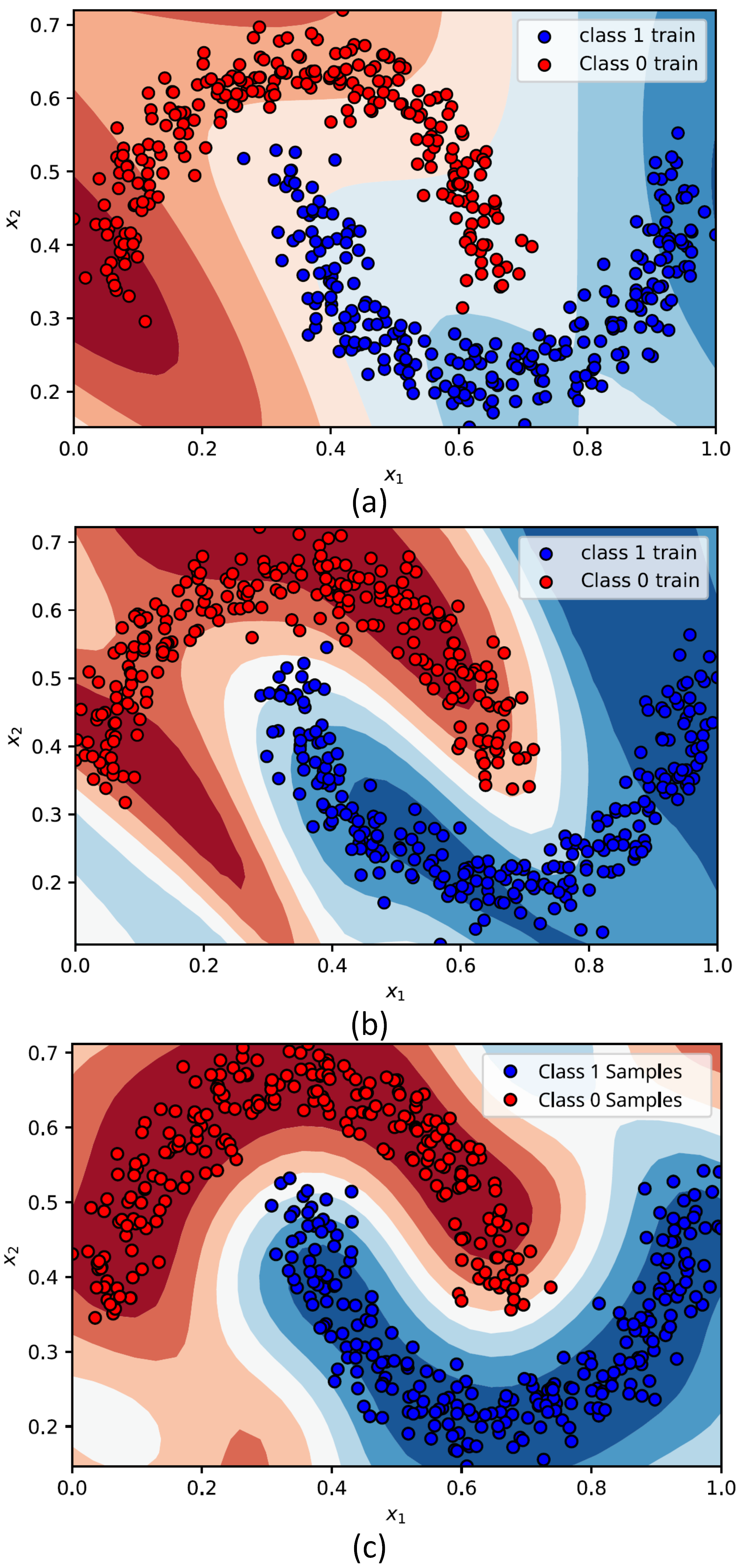}
    \caption{The discriminator network output realized using multiple stages of the trainable encoding circuit $\rm E_1$ (Fig. \ref{fig:quantum_circuits}) and trained using the two-moon dataset. (a) single-stage network, (b) four-stage network, (c) five-stage network.}
    \label{fig:disc_two_moon}
\end{figure}

In Fig. \ref{fig:disc_two_moon}, output of the discriminator network trained using the two-moon dataset shown. The network is realized using multiple stages of the trainable encoding circuit shown in Fig. \ref{fig:quantum_circuits}(a). Increasing the number of stages improves the effectiveness of the boundaries separating the two classes of data samples. The five-stage discriminator network as a quantum classifier can efficiently separate data samples of the two classes with accuracy of 100\% for the class 0 samples and 99\% for the class 1 samples. 

\section{Implementation Results of Hybrid GAN}
The quantum circuits implementation and simulations are performed using the IBM's Qiskit open-source software development kit (SDK) which provides access to prototype superconducting quantum processors through cloud-based quantum computing services.
The training procedure for the hybrid quantum-classical GAN is outlined in Table \ref{table:gan_training}. The discriminator and generator networks are trained iteratively using the mini-batch SGD algorithm with a learning rate of 0.01 and the mini-batch size of 10.

\begin{table}[!t]
\caption{}
\centering
\begin{tabular}{|p{0.9\columnwidth}|}
\hline
\vspace{1pt}
\begin{center}
\textbf{Training Procedure of the Hybrid Quantum-Classical GAN}
\end{center}\\
\hline
\vspace{1pt}
\textbf{Algorithm:} Minibatch Stochastic Gradient Descent 
\vspace{3pt}\\
\hline
\vspace{1pt}
\textbf{Trainable Parameters:} $\boldsymbol{\theta_D}$ and $\boldsymbol{\theta_G}$
\vspace{3pt}\\
\hline
\vspace{1pt}
\textbf{Hyperparameters:} Learning rates $\alpha_D$, $\alpha_G$ and batch size $m$
\vspace{3pt}\\
\hline
\vspace{1pt}

\textbf{for} number of iterations:\\
\setlength\parindent{15pt}
\textbf{for} minibatch steps:\\
\setlength \parindent{30pt} \textbf{Discriminator} \\
\setlength\parindent{30pt}
- Sample subset of length $m$ from dataset.\\
\setlength\parindent{30pt}
- Record $m$ generated samples.\\
\setlength\parindent{30pt}
- Combine sampled with generated samples and assign \\ \setlength \parindent{36pt} correct class labels.\\
\setlength\parindent{30pt}
- Compute loss and discriminator parameter gradients \\ \setlength \parindent{36pt} across all minibatch samples.\\
\setlength\parindent{30pt}
- Update discriminator trainable parameters:
$$\boldsymbol{\theta_{D}^{(k+1)}} = \boldsymbol{\theta_D^{(k)}} - \alpha_D  \frac{1}{m} \sum_{i = 1}^{m} \boldsymbol{\nabla} \mathcal{L}_{D,i} $$
\\
\setlength \parindent{30pt} \textbf{Generator} \\
\setlength \parindent{30pt} - Set discriminator for generator training to updated \\ \setlength\parindent{36pt} discriminator model \\
\setlength \parindent{30pt} - Generate $m$ minibatch samples.\\
\setlength \parindent{30pt} - Compute loss and generator parameter gradients across \\ \setlength\parindent{36pt} all minibatch samples.\\
\setlength\parindent{30pt}
- Update discriminator trainable parameters:
$$\boldsymbol{\theta_{G}^{(k+1)}} = \boldsymbol{\theta_G^{(k)}} - \alpha_G \frac{1}{m}  \sum_{i = 1}^{m} \boldsymbol{\nabla} \mathcal{L}_{G,i} $$
\\
\hline
\end{tabular} \newline
{\label{table:gan_training}.}
\end{table}

\subsection{Uniform Data Distribution}

\begin{figure}
    \centering
    \includegraphics[width=0.9\columnwidth]{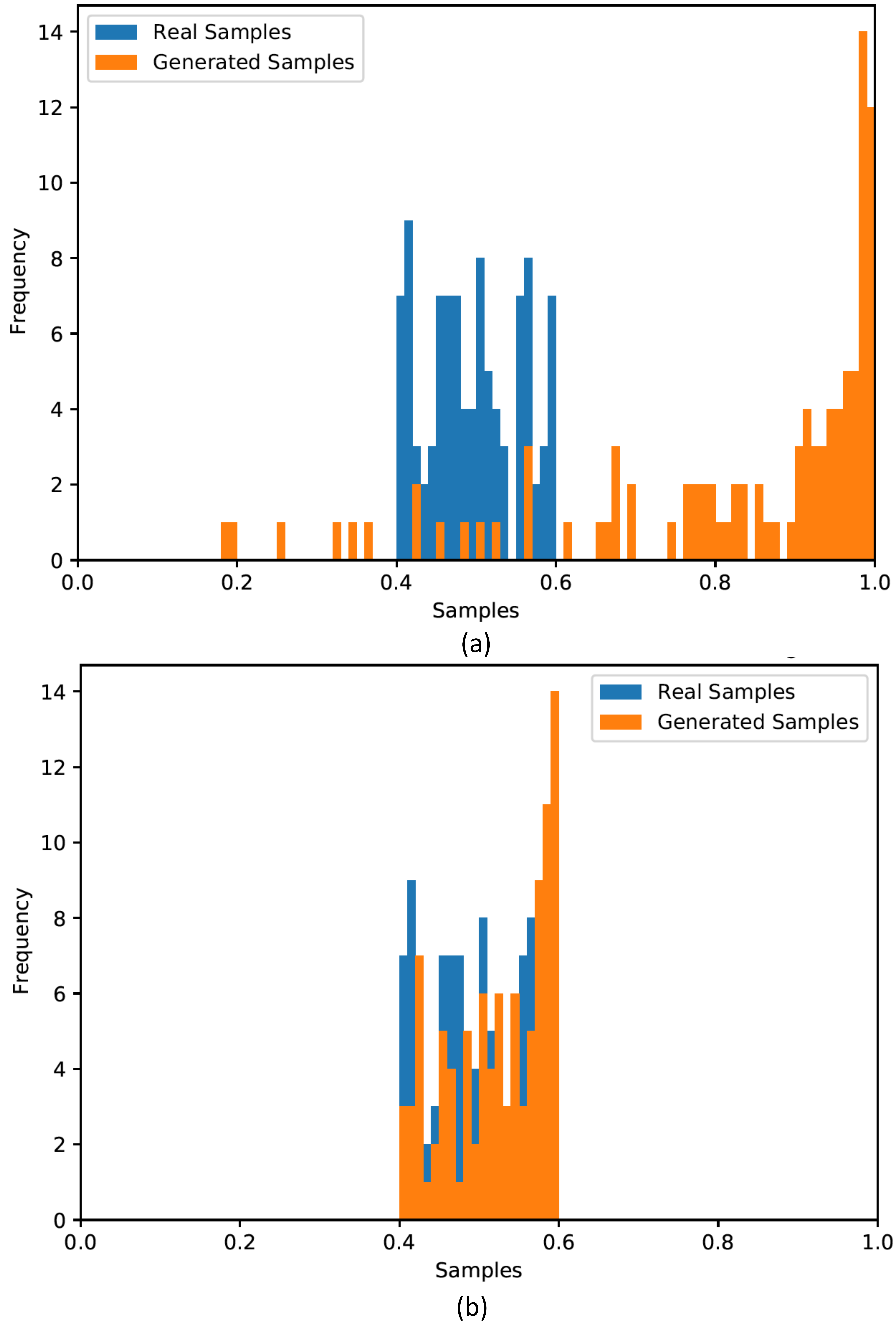}
    \caption{Real and generated data distributions for the hybrid GAN trained using a uniform data distribution. (a) before training the hybrid GAN, (b) after training the hybrid GAN.}
    \label{fig:uniform_training}
\end{figure}

The real (train) and generated data distributions before training the hybrid GAN are shown in Fig. \ref{fig:uniform_training}(a). The generated data samples are achieved using the randomly initiated generator network. The generated data sampled are bounded between 0 and 1. The real data samples are uniformly distributed between 0.4 and 0.6. As expected, there is a very low similarity between the two random distributions. 
In Fig. \ref{fig:uniform_training}(b), the real and generated data distributions after only 60 training epochs of the hybrid GAN are shown. The distribution of the generated data has been significantly changed toward the distribution of the real data. This confirms successful training of the hybrid GAN. The \textit{Inception Score} is a performance metric proposed for classical GAN \cite{salimans2016improved} which partly measures diversity of generated data. However, it is difficult to apply this metric to the hybrid GAN due to the need for large number of data samples (typically 30,000) \cite{salimans2016improved}. 

\begin{figure}
    \centering
    \includegraphics[width=0.9\columnwidth]{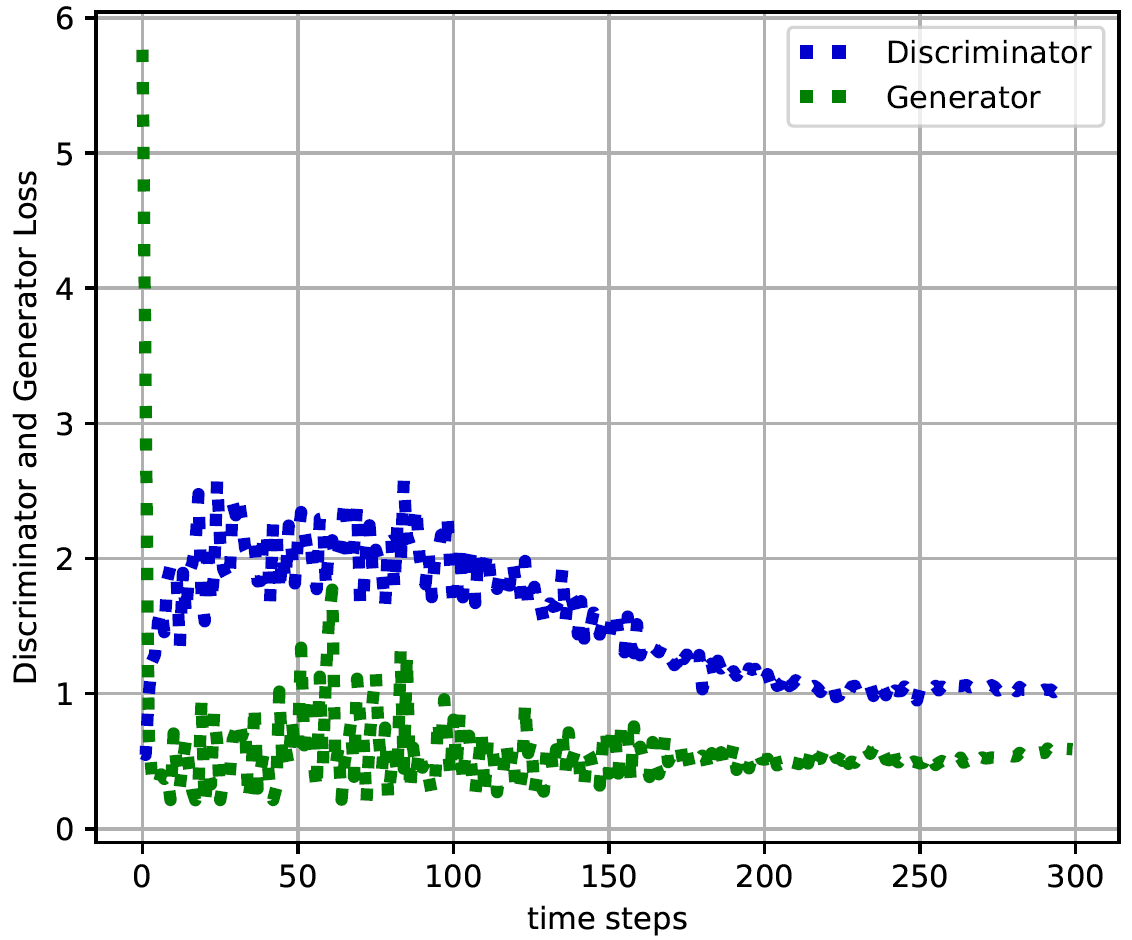}
    \caption{Loss of the discriminator and generator networks for 300 training epochs of the hybrid GAN.}
    \label{fig:loss_disc_gen}
\end{figure}

In Fig. \ref{fig:loss_disc_gen}, loss of the discriminator and generator networks for 300 training epochs of the hybrid GAN is shown. The generator loss is high in the beginning of training as a result of random initialization. As the training proceeds, the discriminator loss increases as the discriminator network is trained to recognize the real data from the generator data. Ultimately, the discriminator and generator loss reach a steady state. In this condition, the generator is effectively able to fool the discriminator, while the discriminator identifies about half of the data samples as the real data and the another half as the fake data. 

\begin{figure}
    \centering
    \includegraphics[width=0.9\columnwidth]{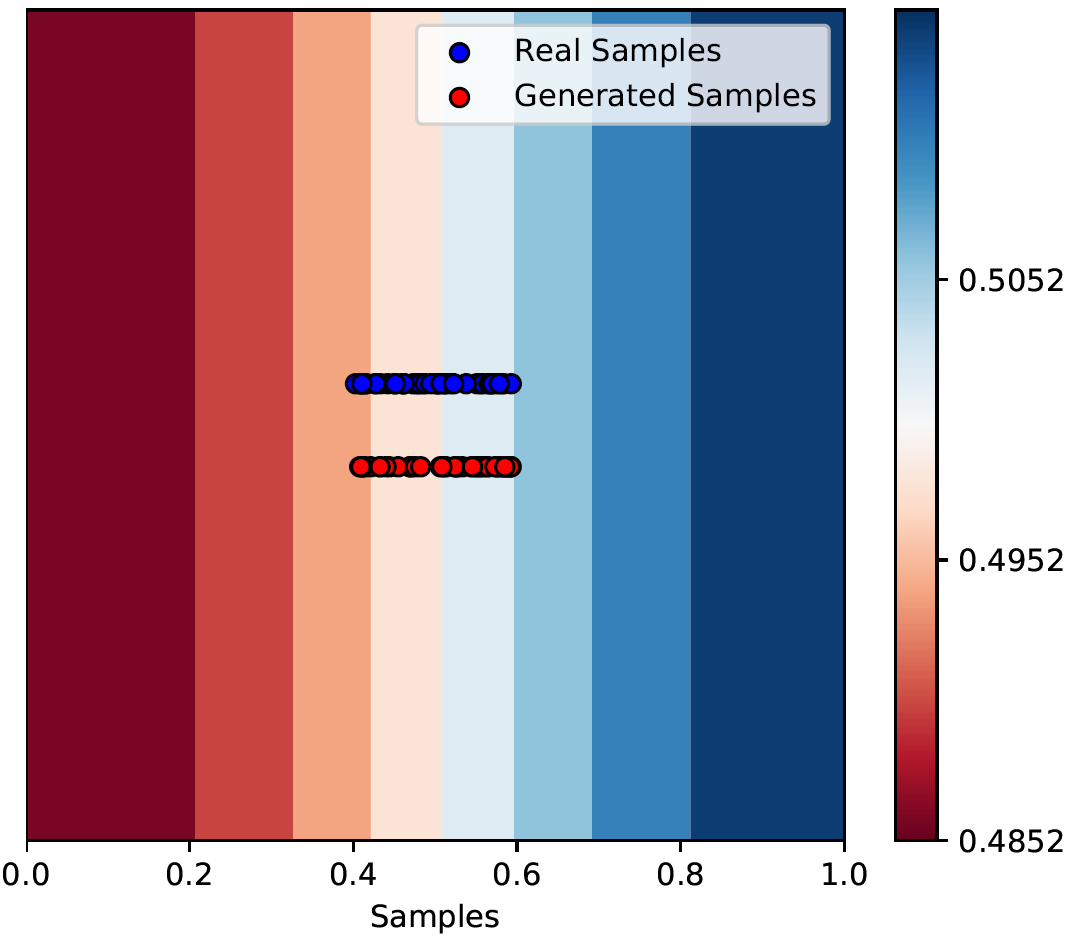}
    \caption{Class boundaries of the discriminator and separated data samples by the hybrid GAN.}
    \label{fig:class_boundaries_gan}
\end{figure}

The class boundaries of the discriminator network and separated data samples by the trained hybrid GAN are shown in Fig \ref{fig:class_boundaries_gan}. The hybrid GAN can effectively separate the real and generated data samples with the boundary close to 0.5.

\subsection{Nonuniform Data Distribution}

\begin{figure}[!t]
\includegraphics[width=0.9\columnwidth]{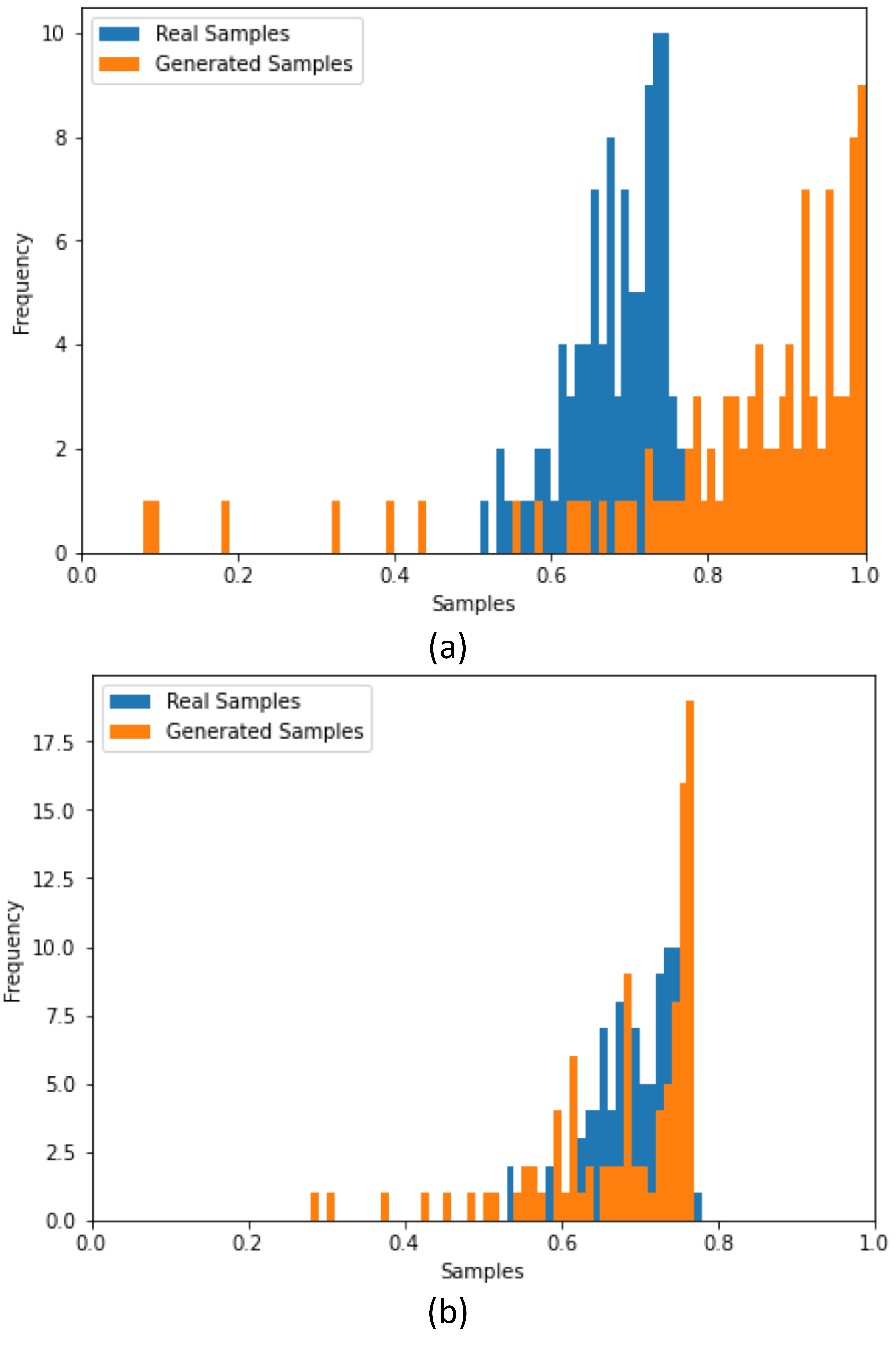}
\caption{Real and generated data distributions for the hybrid GAN trained using a nonuniform data distribution. (a) before training the hybrid GAN, (b) after training the hybrid GAN.}
\label{fig:nonuniform_training} 
\end{figure}

We also evaluate performance of the hybrid GAN using training data samples with a nonuniform distribution. The training data distribution and output data distribution of the generator with random initial states are shown in Fig. \ref{fig:nonuniform_training}(a). There is a very low similarity between the distributions. The real and generated data distributions after training the hybrid GAN are shown in Fig. \ref{fig:nonuniform_training}(b). The generated data distribution has significantly changed toward the real data distribution. We evaluate similarity between the real and generated data distributions using the Kullback-Leibler (KL) and the Jensen–Shannon (JS) divergence scores 
\begin{equation}
    \label{KL_score}
    KL (p_d||p_g) = \sum_{i} p_d(x_i) \log \left( \frac{p_d(x_i)}{p_g(x_i)} \right)
\end{equation}
\begin{equation}
    \label{JS_score}
    JS (p_d||p_g) = \frac{1}{2} KL \left( p_d \Big{|}\Big{|} \frac{p_d + p_g}{2} \right) + \frac{1}{2} KL \left( p_g \Big{|}\Big{|} \frac{p_d + p_g}{2} \right)
\end{equation}
where $p_d(\boldsymbol{x})$ is the probability density of real data $\boldsymbol{x}$ and $p_g(\boldsymbol{x})$ is the posterior probability density of generated data $G(\boldsymbol{z})$. In the optimum condition $p_d(\boldsymbol{x}) = p_g(\boldsymbol{x})$, the KL and JS scores will be zero. 
For the data distributions in Fig. \ref{fig:nonuniform_training}, the KL score is 1.71 before training and 0.39 after training the hybrid GAN. The JS score decreases from 2.63 before training to 0.52 after training. These results validate successful operation of the hybrid GAN for the nonuniform data distribution.

For the nonuniform training data distribution, unlike the uniform distribution, the discriminator network features different certainty in separating the data samples as real/fake across the range of samples. This mitigates a training failure in the hybrid GAN namely the mode collapse, which will be discussed in Section V-C.

\subsection{Training Challenges of Hybrid GAN}

The training process of the hybrid GAN has been quite challenging and has encountered multiple issues including convergence failure, slow training, and mode collapse. 

\subsubsection{Convergence Failure}
Convergence of the GAN, in either classical or quantum domain, is difficult as a result of the simultaneous training of the generator and discriminator networks in a zero-sum game. In every training epoch that parameters of one of the models are updated, the optimization problem for the other model is changed. This means that an improvement in one model can lead to deterioration of the other model. This trend can be observed in Fig. \ref{fig:loss_disc_gen} epochs 1 to 100 where the initial improvements in the generator loss lead to increasing loss of the discriminator. The vanishing gradient problem in the GAN can be due to an over-trained discriminator which rejects most of the generator samples (rather than half of the samples). This issue can be prevented by avoiding over-training of the discriminator network \cite{GAN, salimans2016improved}. If the GAN training is successful, the two networks reach a state of equilibrium between the two competing loss criteria. This phase of the training can be noted in Fig. \ref{fig:loss_disc_gen} epochs 100 to 200. Afterward, losses of the generator and discriminator networks are plateaued, indicating that the training process has converged. 

\subsubsection{Slow Training}
The barren plateaus as a general problem in the training of QNNs was discussed in Section II-D. The barren plateaus is a vanishing gradients issue which can extremely slow down or stop the progress of training process. It can be mitigated through modified initial state of the quantum circuits and, if not successful, changing the quantum circuits architecture. However, a major challenge is that it cannot be predicted if a certain initial state or a network architecture will encounter this issue before the network is trained for a sufficient number of iterations which itself is also unknown. As a result, the development and training of the QNNs usually entail several trials and revisions of the network architecture. 

\subsubsection{Mode Collapse}
\begin{figure}[!t]
\includegraphics[width=0.9\columnwidth]{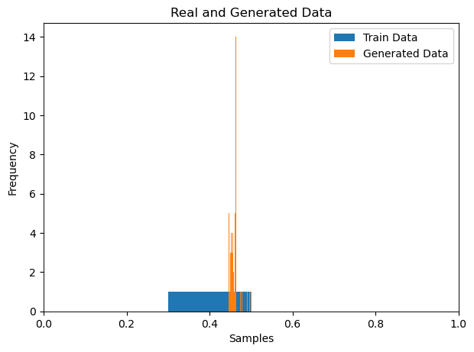}
\caption{\label{fig:mode_collapse} Distributions of the real and generated data samples for the hybrid GAN with mode collapse failure.}
\end{figure}

The convergence of training process is not a sufficient condition to guarantee successful realization of the GAN. Another important issue is the \textit{mode collapse} which refers to the condition that the generator learns to produce only a limited range of samples from the real data distribution. We have encountered the mode collapse during training of the hybrid GAN. In Fig. \ref{fig:mode_collapse}, distributions of the real and generated data samples are shown in the case of hybrid GAN with mode collapse failure. The generated data samples are concentrated around a limited range of samples (close to 0.45), while for most of the real data samples there is no corresponding generated data samples. We have resolved this issue by using different quantum circuits to realize the generator QNN. 

Furthermore, we have noted that the mode collapse occurs more frequently when the GAN is trained using a uniform distribution. This is a result of the equal probability of the data samples which the discriminator network classifies them as the real data. For the nonuniform training data distribution, however, the certainty of the discriminator in real/fake data classification is different across the data samples.

In the classic GANs, some modified methods are developed to extend the range of generated data samples, including Wasserstein GAN \cite{arjovsky2017wasserstein} and Unrolled GAN \cite{metz2016unrolled}. It is an opportunity for future research to evaluate effectiveness of such approaches in the hybrid quantum-classical GAN.

\section{Conclusion}
We presented a hybrid quantum-classical generative adversarial network (GAN). The generator quantum neural network (QNN) is realized using a fixed encoding circuit followed by a parameterized quantum circuit. The discriminator QNN is implemented using a trainable encoding circuit through a modular design approach which enables embedding of classical data into a quantum circuit without using explicit encoding and ansatz circuits. It is shown that gradient of the discriminator and generator loss functions can be calculated using the same quantum circuits which have realized the discriminator and generator QNNs. The developed hybrid quantum-classical GAN is trained successfully using uniform and nonuniform data distributions. Using the nonuniform distribution for training data, the mode collapse failure, which GANs are prone to, can be mitigated. The proposed approach for the realization of hybrid quantum-classical GAN can open up a research direction for the implementation of more advanced GANs on the near-term quantum processors.

\bibliographystyle{IEEEtran}
\bibliography{ref}

\end{document}